\documentclass[aps,preprint]{revtex4}
\usepackage{amsmath}
\usepackage[english]{babel}
\usepackage[latin1]{inputenc}
\usepackage{amssymb}
\usepackage{graphicx}
\usepackage{natbib}
\usepackage{epsfig}
\usepackage{bm}
\usepackage{hyperref}
\usepackage{dcolumn}
\usepackage{graphicx}
\usepackage{color}

\begin{document}

\title{\sc\Large{Weak decays of magnetized charged pions in the symmetric gauge}}

\author{M. Coppola$^{a,b}$, D. Gomez Dumm$^{c}$, S. Noguera$^{d}$ and N.N.\ Scoccola$^{a,b}$}

\affiliation{$^{a}$ CONICET, Rivadavia 1917, 1033 Buenos Aires, Argentina}
\affiliation{$^{b}$ Physics Department, Comisi\'{o}n Nacional de Energ\'{\i}a At\'{o}mica, }
\affiliation{Av.\ Libertador 8250, 1429 Buenos Aires, Argentina}
\affiliation{$^{c}$ IFLP, CONICET $-$ Departamento de F\'{\i}sica, Fac.\ de Cs.\ Exactas,
Universidad Nacional de La Plata, C.C. 67, 1900 La Plata, Argentina}
\affiliation{$^{d}$ Departamento de F\'{\i}sica Te\'{o}rica and IFIC, Centro Mixto
Universidad de Valencia-CSIC, E-46100 Burjassot (Valencia), Spain \vspace*{2cm}}

\begin{abstract}
We consider the decay $\pi^- \to l \, \bar{\nu}_l$ ($l=e^-,\,\mu^-$) in
the presence of an arbitrary large uniform magnetic field, using the
symmetric gauge. The consequences of the axial symmetry of the problem and
the issue of angular momentum conservation are discussed in detail. In
particular, we analyze the projection of both the canonical and the
mechanical total angular momenta along the direction of the magnetic field.
It is found that while the former is conserved in the symmetric gauge, the
latter is not conserved in both the symmetric and Landau gauges. We derive
an expression for the integrated $\pi^- \to l \, \bar{\nu}_l$ width that
coincides exactly with the one we previously found using the Landau gauge,
providing an explicit test of the gauge
independence of that result. Such an expression implies that for nonzero
magnetic fields the decay width does not vanish in the limit in which the
outgoing charged leptons can be considered as massless, i.e.\ it does not exhibit the
helicity suppression found in the case of no external field.
\end{abstract}
%\date{\today}

\pacs{}

\maketitle

\renewcommand{\thefootnote}{\arabic{footnote}}
\setcounter{footnote}{0}

\section{Introduction}

Recently, a significant interest has been devoted to the effect of intense
magnetic fields on the properties of strongly interacting
matter~\cite{Kharzeev:2012ph,Andersen:2014xxa,Miransky:2015ava}. This is
mostly motivated by the realization that strong magnetic fields might play
an important role in the study of the early Universe~\cite{Grasso:2000wj},
in the analysis of high energy noncentral heavy ion collisions~\cite{HIC},
and in the physics of stellar objects like the magnetars~\cite{duncan}. It
is well known that magnetic fields induce interesting phenomena, such as the
enhancement of the QCD vacuum (the so-called ``magnetic
catalysis'')~\cite{Gusynin:1994re} and the decrease of critical temperatures
for chiral restoration and deconfinement QCD transitions~\cite{Bali:2011qj}.
More recently, it has also been shown that an external magnetic field can
lead to a significant enhancement of the leptonic decay widths of charged
pions~\cite{Bali:2018sey,Coppola:2019idh}. It should be noticed
that the hadronic decay rates in external magnetic fields are important for the
stability and equilibrium of magnetars. Moreover,
their dominant cooling mechanisms involve (inverse) $\beta$-decay, photo-meson interactions
and pion decay \cite{Waxman:1997ti}. Pions radiate energy via inverse Compton scattering
until they decay, imprinting the spectrum of the subsequently produced neutrinos
\cite{Zhang:2002xv}.
Although several studies of weak decays of hadrons under strong
electromagnetic fields have been reported in the
literature~\cite{Nikishov:1964zza,Nikishov:1964zz,Matese:1969zz,FassioCanuto:1970wk},
in most works the effects of the external fields on the relevant hadronic matrix elements
have not been fully considered. In the case of charged pions,
these effects have been recently analyzed in the context of chiral
perturbation theory~\cite{Andersen:2012zc}, effective chiral
models~\cite{Simonov:2015xta, Liu:2018zag, Coppola:2018vkw,Coppola:2019uyr}
and lattice QCD (LQCD) calculations~\cite{Bali:2018sey}.
The most general form of the relevant hadronic matrix elements for the
processes $\pi^-\to l \, \bar\nu_l$ ($l=e^-,\,\mu^-$) has been obtained through a
model-independent analysis in Ref.~\cite{Coppola:2018ygv}, where the effect
of a magnetic field $\vec B$ on both pion and lepton wave functions is fully
taken into account. In particular, it is found that the vector and axial
vector pion-to-vacuum matrix elements can be parametrized in general through
one and three hadronic form factors, respectively. Taking into account the
expression for the $\pi^-\to l \,\bar\nu_l$ decay width in
Ref.~\cite{Coppola:2018ygv}, quantitative predictions have been given in
Ref.~\cite{Coppola:2019idh} for magnetic fields up to $eB\simeq 1$~GeV$^2$.
Those results, based on the estimations given in Ref.~\cite{Coppola:2019uyr}
for the hadronic form factors and the pion mass within an effective Nambu--Jona-Lasinio
model, show a strong enhancement of the total width with respect to its
value for $B=0$. In addition, it is seen that the presence of the magnetic
field affects dramatically the ratio between muonic and electronic decay
rates~\cite{Coppola:2019idh}. This is related to the fact that the widths do
not vanish in the limit of vanishing lepton masses, as happens to be the
case for $B=0$.

For $\pi^- \to l \,\bar\nu_l$ decays, the presence of an external magnetic
field has also an important effect on the angular distribution of outgoing
antineutrinos. In Ref.~\cite{Bali:2018sey}, on the basis of some assumptions
related to angular momentum conservation and chirality, it is claimed that
the momentum $\vec k$ of the antineutrino has to be parallel to the magnetic
field (i.e., the perpendicular component $\vec k_\perp$ has to be zero).
This is in contradiction with the analysis in Ref.~\cite{Coppola:2019idh},
where it is found that $\vec k_\perp$ is in fact the dominant component of
the momentum for magnetic fields larger than about 0.1~GeV$^2$. According to
Ref.~\cite{Coppola:2018ygv}, conservation laws do not imply $\vec k_\perp
=0$, therefore one should integrate over all possible values of the
antineutrino momentum. At this point it should be noticed that in
Ref.~\cite{Coppola:2018ygv} the calculation of the general form of the
$\pi^-\to l \, \bar\nu_l$ decay width has been carried out using expressions
for the charged pion and lepton wave functions in the Landau gauge. Although
this is in fact the most widely chosen gauge to perform this type of
calculations, it may be not the most convenient one when dealing with
arguments of angular momentum conservation, such as those considered in
Ref.~\cite{Bali:2018sey}. As noticed in Ref.~\cite{Wakamatsu:2017isl}, the
consequences of the axial symmetry of the problem, as well as the physical
meaning of angular momenta, can be better understood if one works in the
symmetric gauge. Having this in mind, the purpose of the present work is to
rederive the expression for the $\pi^-\to l \,\bar\nu_l$ decay width
including the most general hadronic matrix elements in the presence of a
uniform magnetic field, now considering the symmetric gauge. It is shown
that this procedure enables a detailed discussion of angular momentum
conservation issues in decay processes of magnetized charged particles. In
addition, our calculation is found to confirm the expected gauge
independence of the general expression for the decay width.

This paper is organized as follows. In Sec.~II we present the expressions of
the different charged fields used in our work and discuss in detail angular
momentum and chirality properties. In Sec.~III we obtain explicit
expressions for the $\pi^\pm$ leptonic weak decay amplitudes. We consider
both the general case of an arbitrary magnetic field strength and the limit
of strong fields. Then we obtain the expression for the decay width by
summing and integrating over all possible final states. Our main conclusions
are presented in Sec.~IV. We also include Appendixes A, B, C and D to quote
some technical details of our calculations.

\section{Preliminaries: Charged particle fields under a uniform magnetic field in the
symmetric gauge}

Let us start by quoting the expressions for the different charged fields
considered in our work, written in terms of particle creation and
annihilation operators. We use the Minkowski metric $g^{\mu\nu} =
\mbox{diag}(1,-1,-1,-1)$, while for a space-time coordinate four-vector
$x^\mu$ we adopt the notation $x^\mu = (t, \vec r\,)$, with $\vec r
=(r_x,r_y,r_z)$. Assuming the presence of a uniform magnetic field $\vec B$,
we orientate the spatial axes in such a way that $\vec B = B\ \hat r_z$, and
consider the symmetric gauge, in which ${\cal A}^\mu = (0,\vec A)$ with
$\vec A = \vec r \times \vec B/2 = (-B\, r_y /2, B \, r_x /2 , 0)$. The
expressions for the charged particle fields can be obtained using the method
described e.g.~in Sec.~19 of Ref.~\cite{Sokolov:1986nk}. For the reader's
convenience we quote here these expressions explicitly, since they are not
commonly given in the literature (in comparison to those corresponding to
the Landau gauge). In the last subsection we discuss some issues related to
the quantum numbers of particle states.

\subsection{Charged pion field}

The charged pion fields can be written as
\begin{equation}
\phi^s_{\pi^{\sigma}}(x) \ = \ \phi^s_{\pi^{-\sigma}}(x)^\dagger \ = \
\sum_{\ell,\imath=0}^\infty \int \frac{ dp_z}{(2\pi)^3\, 2 E_{\pi^{\sigma}}}
\left[
a^\sigma(\breve p)  \; W^s_{\bar{p}}(x) + a^{-\sigma}(\breve p)^\dagger  \; W^{-s}_{\bar{p}}(x)^\ast \right]\ ,
\label{chargepionexp}
\end{equation}
where  $Q_{\pi^\sigma}=\sigma |e|$ is the pion charge with $\sigma=\pm$, $s
= \mbox{sign}(Q_{\pi^\sigma} B)$ and $B_e=|Q_{\pi^\sigma} B|=|eB|$. Note
that if $B>0$ then $s=\sigma$. The pion energy is given by $E_{\pi^{\sigma}}
= \sqrt{m^2_{\pi^{\sigma}}+ (2\ell+1) B_e + p_z^2}$. We have also defined
$\bar p = (E_{\pi^{\sigma}},\breve{p})$ and $\breve{p} = (\ell,\imath,p_z)$, where $p_z$ is an arbitrary real number
while $\ell$ and $\imath$ are non-negative integer numbers. The functions $W^s_{\bar{p}}(x)$ are solutions of the eigenvalue equation
\begin{equation}
{\cal D}_\mu {\cal D}^\mu \  W^s_{\bar{p}}(x) \ = \ -  \left[ E_{\pi^{\sigma}}^2 - (2 \ell+1) B_e - p_z^2 \right]  W^s_{\bar{p}}(x)\ ,
\label{ecautovbo}
\end{equation}
where ${\cal D}^\mu = \partial^\mu + i s B_e \, ( r_x \, \delta^{\mu 2} -
r_y \, \delta^{\mu 1})/2 $. Introducing polar coordinates $\rho,\phi$ in the
plane perpendicular to the magnetic field, their explicit form is given by
\begin{equation}
 W^s_{\bar{p}}(x) \ = \
\sqrt{2\pi} \ e^{-i(E_{\pi^{\sigma}} t - p_z r_z)}\, e^{-i s (\ell - \imath) \phi} \, R_{\ell,\imath}(\rho)\ ,
\label{efes}
\end{equation}
where
\begin{equation}
R_{\ell,\imath}(\rho) \ = \ N_{\ell,\imath} \ \xi^{(\ell - \imath)/2} \  e^{-\xi/2} \ L_\imath^{\ell-\imath}(\xi)\ .
\end{equation}
Here we have used the definitions $N_{\ell, \imath} = (B_e \ \imath! /
\ell!)^{1/2}$ and $\xi = B_e \, \rho^2/2\,$, while $L_m^\alpha(x)$ are the
associated Laguerre polynomials. It can be seen that the functions
$W^s_{\bar{p}}(x) $ satisfy the orthogonality relations
\begin{equation}
\int d^3r \, W^s_{\bar{p}\,'}(x)^\ast \ W^s_{\bar{p}}(x) \ = \
(2\pi)^3 \,\delta_{\ell\ell'} \,\delta_{\imath \imath'} \,\delta(p_z-p_z^{\,\prime})\  . \label{orthF}
\end{equation}
In addition, the creation and annihilation operators in Eq.~(\ref{chargepionexp}) satisfy the commutation relations
\begin{align}
\left[a^\sigma(\breve p) , a^{\sigma '}(\breve p\,')^\dagger\right]  &=
\, 2 E_{\pi^{\sigma}}  \,(2\pi)^3  \, \delta_{\ell\ell'}\, \delta_{\imath \imath'}
\, \delta(p_z-p_z^{\,\prime})\, \delta_{\sigma\sigma'}\ , \nonumber \\
\left[a^\sigma(\breve p), a^{\sigma'}(\breve p\,') \right] &= \, 0\ .
\label{conbos}
\end{align}
Note that with these conventions the operators $a^\sigma(\breve p)$  turn out to have different dimensions from the creation and annihilation
operators that are usually defined in absence of the external magnetic field.

It is also useful to calculate the particle number associated with the state
$|\pi^\sigma(\breve p) \rangle = {a^\sigma}(\breve p)^\dagger |0\rangle$ in
a volume $V$. Given our gauge choice, it is convenient to consider an
infinite area in the $xy$ plane and a finite length $L$ along the $z$ axis.
We obtain
\begin{equation}
n_{\pi^\sigma} \ = \ \int_V  d^3r\;
\langle \pi^\sigma(\breve p) |  \ i \left[ \phi^s_{\pi^\sigma}\!^\dagger \, \partial_t \phi^s_{\pi^\sigma}
- \Big(\partial_t \phi^s_{\pi^\sigma}\!^\dagger\Big)\, \phi^s_{\pi^\sigma} \right] | \pi^\sigma(\breve p)\rangle
\ = \ 2 E_{\pi^{\sigma}} (2\pi)^2 L\ .
\label{numberb}
\end{equation}
Note that we are normalizing to
$8\pi^2 E$ particles per unit length, which differs from the usual
normalization $\rho = n/V=2E$.

\subsection{Charged lepton field}

Assuming the same conventions for the magnetic field and considering the symmetric
gauge, for the charged lepton fields we have
\begin{equation}
\psi_l^s(x) \ = \
\sum_{\tau=1,2} \ \sum_{n,\upsilon=0}^\infty \int \! \frac{dq_z}{(2\pi)^3 \, 2 E_l}
\left[\,
b\left(\breve {q},\tau\right) \, U_l^s\left(x,\breve {q},\tau\right) + d\left(\breve {q},\tau\right)^\dagger
\,  V_l^{-s}\left(x,\breve {q},\tau\right)\,
\right]\ ,
\label{fermionfieldBpart}
\end{equation}
where $\breve {q} = (n,\upsilon,q_z)$, $E_l = \sqrt{m^2_l + 2n B_e + q_z^2}$
and $s = \mbox{sign}(Q_l B)$, with $Q_l=-|e|$ (thus, $B_e =|Q_l B|$). Here,
$q_z$ is an arbitrary real number, while $n$ and $\upsilon$ are non-negative
integer numbers. The creation and annihilation operators satisfy
\begin{alignat}{2}
\left\{b(\breve{q},\tau), b(\breve{q}\,',\tau')^\dagger\right\} &=
\left\{d(\breve{q},\tau), d(\breve{q}\,',\tau')^\dagger\right\} & &=  \,
2 E_l\, (2\pi)^3\, \delta_{\tau \tau'}\, \delta_{nn'}\, \delta_{\upsilon \upsilon'}\, \delta(q_z - q_z')\ , \\
\left\{b(\breve{q},\tau), d(\breve{q}\,',\tau')^\dagger\right\} &=
\left\{d(\breve{q},\tau)^\dagger, b(\breve{q}\,',\tau')^\dagger\right\} & &= \, 0\ .
\label{conferB}
\end{alignat}

For $n>0$ the spinors in Eq.~(\ref{fermionfieldBpart}) are given, in the Weyl basis, by
\begin{eqnarray}
U_l^s\left(x,\breve{q},\tau \right) & = &
\frac{\sqrt{\pi}}{\sqrt{E_l + m_l}} e^{-i (E_l t - q_z r_z)} e^{-i s (n-\upsilon-1/2) \phi}\, \times \nonumber \\
&&\hspace{-2cm}\left[
\delta_{\tau,1} \!
\left( \!\!\! \begin{array}{c}
 e^{-i\phi/2} \ \varepsilon_- \  R_{n_{s+},\upsilon}(\rho) \\
 i s e^{i\phi/2} \ \sqrt{2 n B_e} \  R_{n_{s-},\upsilon}(\rho) \\
 e^{-i\phi/2} \ \varepsilon_+ \ R_{n_{s+},\upsilon}(\rho)\\
-i s e^{i\phi/2} \ \sqrt{2 n B_e} \ R_{n_{s-},\upsilon}(\rho) \\
\end{array}
\!\! \right)
+
\delta_{\tau,2}
\left( \!\!\! \begin{array}{c}
-i s e^{-i\phi/2} \ \sqrt{2 n B_e} \  R_{n_{s+},\upsilon}(\rho) \\
e^{i\phi/2} \ \varepsilon_+    \ R_{n_{s-},\upsilon}(\rho) \\
 i s e^{-i\phi/2} \ \sqrt{2 n B_e} \  R_{n_{s+},\upsilon}(\rho) \\
 e^{i\phi/2} \ \varepsilon_-   \ R_{n_{s-},\upsilon}(\rho)\\
\end{array}
\!\! \right)
\right]\ ,
\end{eqnarray}
\begin{eqnarray}
V_l^{-s}\left(x,\breve{q},\tau\right) & = &
\frac{\sqrt{\pi}}{\sqrt{E_l + m_l}} e^{i (E_l t - q_z r_z)} e^{-i s (n-\upsilon-1/2) \phi}\,\times \nonumber \\
&&\hspace{-2cm} \left[
\delta_{\tau,1} \!
\left( \!\!\! \begin{array}{c}
i s e^{-i\phi/2} \ \sqrt{2 n B_e} \ R_{n_{s+},\upsilon}(\rho) \\
e^{i\phi/2} \ \varepsilon_+    \ R_{n_{s-},\upsilon}(\rho) \\
 i s e^{-i\phi/2} \ \sqrt{2 n B_e} \ R_{n_{s+},\upsilon}(\rho) \\
 - e^{i\phi/2} \ \varepsilon_-   \ R_{n_{s-},\upsilon}(\rho)\\
\end{array}
\!\!\! \right) \! \!
+
\delta_{\tau,2}
\left( \!\! \begin{array}{c}
- e^{-i\phi/2} \ \varepsilon_- \  R_{n_{s+},\upsilon}(\rho) \\
 i s e^{i\phi/2} \ \sqrt{2 n B_e} \ R_{n_{s-},\upsilon}(\rho) \\
 e^{-i\phi/2} \ \varepsilon_+ \ R_{n_{s+},\upsilon}(\rho)\\
i s e^{i\phi/2} \ \sqrt{2 n B_e} \ R_{n_{s-},\upsilon}(\rho) \\
\end{array}
\!\! \right)
\right]\ .
\end{eqnarray}
where $\varepsilon_\pm = E_l + m_l \pm q_z$. Here the non-negative integer index $n_{s\lambda}$ is related to the quantum number $n$ by $n_{s\pm}  \ = \ n - (1\mp s)/2$.

In the particular case of the lowest Landau level (LLL) $n=0$, only one
polarization is allowed. Using
the notation $\breve q_{LLL} = (0,\upsilon,p_z)$, the explicit form of the
spinors in this case are
\begin{align}
U_l^s\left(x,\breve q_{LLL}\right) &= \frac{\sqrt{\pi}}{\sqrt{E_l+m_l}} e^{-i (E_l t - q_z r_z)} e^{i s \upsilon \phi} R_{0,\upsilon}(\rho)
\left[ \delta_{s,1} \ \left( \!\!\! \begin{array}{c}
\ \varepsilon_- \ \ \\
0 \\
\varepsilon_+\\
0 \\
\end{array}
\!\!\! \right) + \,\delta_{s,-1} \left( \!\!\! \begin{array}{c}
0 \\
\ \varepsilon_+ \ \ \\
0 \\
\varepsilon_-\\
\end{array}
\!\!\! \right)\ \right]\ ,
\label{ULLL}  \\
V_l^{-s}\left(x,\breve q_{LLL}\right) &=
\frac{\sqrt{\pi}}{\sqrt{E_l+m_l}} e^{i (E_l t - q_z r_z)} e^{i s \upsilon \phi} R_{0,\upsilon}(\rho)
%\nonumber \\
%&&
%\!\!\!\!\!\!\!\!\!
 \left[ \delta_{s,1} \ \left( \!\!\!
\begin{array}{c}
-\varepsilon_-\\
0 \\
\varepsilon_+\\
0 \\
\end{array}
\!\!\! \right) + \,\delta_{s,-1}  \  \left( \!\!\!
\begin{array}{c}
0 \\
\varepsilon_+\\
0 \\
-\varepsilon_-\\
\end{array}
\!\!\! \right)\ \right]\ . \label{VLLL}
\end{align}

It can be seen that the spinors satisfy the orthogonality relations
\begin{alignat}{2}
\int\! d^3 r  \, U^s(x,\breve{q}, \tau)^\dagger U^s(x,\breve{q}\,'\!, \tau') &=
\hphantom{-} \int\! d^3 r \, V^{-s}(x,\breve{q}, \tau)^\dagger V^{-s}(x,\breve{q}\,'\!, \tau') & &=
2 E_l  \, (2\pi)^3 \delta_{\breve{q},\breve{q}^{\prime}} \, \delta_{\tau\tau'}\ ,
\nonumber \\
\int\! d^3 r  \, U^s(x,\breve{q}, \tau)^\dagger V^{-s}(x,\breve{q}\,'\!, \tau') &=
\hphantom{-} \int\! d^3 r \, V^{-s}(x,\breve{q}, \tau)^\dagger {U^s}(x,\breve{q}\,'\!, \tau') & &= 0\ , \nonumber \\
\int\! d^3 r  \, \bar U^s(x,\breve{q}, \tau) U^s(x,\breve{q}\,'\!, \tau') &=
- \int\! d^3 r \, \bar V^{-s}(x,\breve{q}, \tau) {V^{-s}}(x,\breve{q}\,'\!, \tau') & &=  2 m_l  \, (2\pi)^3
\delta_{\breve{q},\breve{q}^{\prime}} \, \delta_{\tau\tau'}\ ,
\nonumber \\
\int\! d^3 r  \, \bar U^s(x,\breve{q}, \tau) V^{-s}(\tilde x,\breve{q}\,'\!, \tau')  &=
\hphantom{-} \int\! d^3 r \, \bar V^{-s}(x,\breve{q}, \tau) U^s(\tilde x,\breve{q}\,'\!, \tau') & &= 0\ ,
\label{cap}
\end{alignat}
which are valid for both $n=0$ and $n>0$. We have used here the definitions
$\delta_{\breve{q},\breve{q}^{\prime}}=\delta_{nn'}\delta_{\upsilon
\upsilon'} \delta\left(q_z-q_z^{\prime}\right)$ and $\tilde x = (t,- \vec
r\,)$.

An alternative representation of the spinors in
Eq.~(\ref{fermionfieldBpart}), closer to the Ritus representation often used
in the Landau gauge, is given in Appendix A.

\subsection{Commutation relations and quantum numbers in the symmetric gauge}

In this subsection we discuss some properties of the operators and particle
states. We consider first the case of charged leptons. We recall that,
in this case, if a physical quantity has an associated quantum mechanical
operator ${\cal O}$, the field theoretical realization of this operator is
given by
\begin{eqnarray}
\tilde {\cal O} = \int_V d^3r \ : \psi^s_l(x)^\dagger \ {\cal O} \ \psi^s_l(x) : \ .
\end{eqnarray}

Let us recall that the Dirac Hamiltonian for a charged particle in a uniform
magnetic field is given by
\begin{eqnarray}
H = \gamma^0 \left [ \vec \gamma \cdot \vec P + m\right] \ .
\label{Ham}
\end{eqnarray}
Here $\vec P$ is the {\it mechanical} momentum, related to the {\it
canonical} momentum $\vec p = -i \; \vec \nabla$ by
\begin{eqnarray}
\vec P = \vec p - Q_l \ \vec A\ ,
\end{eqnarray}
where $\vec A$ is the vector potential associated to the uniform magnetic
field $\vec B$. Although the explicit relation between both momenta depends
on the gauge choice, it is seen that $\vec P$ is a gauge covariant quantity. For a
magnetic field orientated along the $z$ direction, the relation
\begin{eqnarray}
[\tilde P_j, \tilde P_k] = i  B \left(\delta_{j1} \delta_{k2} - \delta_{j2}
\delta_{k1}\right) \tilde Q
\end{eqnarray}
is found to be satisfied (integer indices $1,2,3$ are intended to be
equivalent to $x,y,z$). Here,
\begin{eqnarray}
\tilde Q = Q_l \; \sum_{\tau=1,2} \ \sum_{n,\upsilon=0}^\infty \int \! \frac{dq_z}{(2\pi)^3}
\left[ b(\breve q,\tau)^\dagger  b(\breve q,\tau) - d(\breve q,\tau)^\dagger  d(\breve q,\tau) \right]\ .
\end{eqnarray}

Using the spinors defined in the previous
subsection, a straightforward calculation shows that, as expected, in the
symmetric gauge one gets
\begin{eqnarray}
\tilde H |l(\breve q, \tau) \rangle = E_l \, |l(\breve q, \tau) \rangle\ ,
\qquad\qquad \tilde H |\bar l(\breve q, \tau) \rangle = E_l \, |\bar
l(\breve q, \tau) \rangle\ ,
\end{eqnarray}
where $E_l = \sqrt{m^2_l + 2n B_e + q_z^2}\,$.

We introduce now the {\it canonical} orbital angular momentum operator $\vec
l = \vec r \times \vec p$ and the spin operator $\vec S$. Given the
fact that the magnetic field breaks the rotational invariance, only the
components of these operators along the $z$ axis are relevant. Using
cylindrical coordinates, the $z$ components $l_z$ and $S_z$ are given
by
\begin{eqnarray}
l_z = -i \frac{\partial}{\partial\phi}\ , \qquad\qquad S_z = \frac{1}{2}
\ \mbox{diag}(1,-1,1,-1)\ .
\end{eqnarray}
Defining the {\it canonical} total  angular momentum as $j_z = l_z +
S_z$ and using the spinors defined in the previous
subsection we obtain
\begin{eqnarray}
\tilde j_z |l(\breve q, \tau) \rangle =  j^{(l)}_z \ |l(\breve q, \tau) \rangle\ ,
\qquad \qquad  \tilde j_z |\bar l(\breve q, \tau) \rangle = - j^{(l)}_z \ |\bar l(\breve q, \tau) \rangle\ ,
\end{eqnarray}
with
\begin{eqnarray}
j^{(l)}_z = -s (n - \upsilon -1/2)\ .
\label{jotalep}
\end{eqnarray}
Thus, as expected, we see that for the charged leptons in the symmetric
gauge one can find energy eigenstates that are also eigenstates of the $z$
component of the total {\it canonical} angular momentum. Since the
energy eigenvalues do not depend on $\upsilon$, we see that, in the
symmetric gauge, a state of a given Landau level $n$ is in general a linear
combination of an infinite set of degenerate states with well defined
quantum number $j^{(l)}_z$. This is consistent with the fact that, in this
gauge, one has $\big[ \tilde j_z, \tilde H\big]=0$, as can be verified from
the previously given expressions for $H$ and $j_z$. We stress here that $j_z$
is {\it not} a gauge covariant quantity. In the Landau gauge, for example,
it is not difficult to check that $\tilde j_z$ does not commute with $\tilde H$, therefore
in general it is not expected to be conserved. Turning back to the symmetric
gauge, it is worth noticing that only the {\it canonical} total angular
momentum is well defined, i.e., energy eigenstates are not in general
eigenstates of $\tilde l_z$ and $\tilde S_z$ separately.

Associated to the {\it mechanical} momentum $\vec P$ we can define the {\it
mechanical} orbital angular momentum $\vec L = \vec r \times \vec P$ and the
{\it mechanical} total angular momentum $\vec J = \vec L + \vec  S\,$.
In the same way as $\vec P$, $\vec J$ is a gauge covariant operator. An
explicit calculation shows, however, that for a magnetic field along the
$z$-axis one has
\begin{eqnarray}
\left[\tilde J_z, \tilde H \right] = -i s B_Q \int d^3r\ \bar \psi(x) \left(r_x \, \gamma^1 + r_y \, \gamma^2\right)\psi(x) \ ,
\end{eqnarray}
which is valid in {\it any} gauge. Hence, $J_z$ is a gauge covariant
quantity but it is {\it not} a conserved quantity. In particular, an explicit calculation
in the symmetric gauge shows that $\tilde J_z$ is not diagonal in the basis of
energy eigenstates.

Let us consider now the limit in which the charged lepton mass $m_l$
vanishes. This is interesting when the magnetic field is relatively
strong, say $B_e \gg m_l^2$. In the limit $m_l = 0$ the chirality operator
$\gamma_5$ becomes equivalent to the helicity operator
$\hat P \cdot \vec S\,$, and commutes with $H$. Consequently, one can obtain energy
eigenstates of well defined chirality/helicity. For arbitrary $n>0$ they
can be constructed as linear combinations of the two polarization states. We
get
\begin{eqnarray}
|l (\breve q, L) \rangle_{\rm ch} &=& -i \frac{\sqrt{E_l - q_z}}{\sqrt{2 E_l}} \ |l(\breve q,1)\rangle_{\rm ch} +  \frac{\sqrt{E_l + q_z}}{\sqrt{2 E_l}} \ |l(\breve q,2)\rangle_{\rm ch} \ , \\
|l (\breve q, R) \rangle_{\rm ch} &=&  i \frac{\sqrt{E_l + q_z}}{\sqrt{2 E_l}} \ |l(\breve q,1)\rangle_{\rm ch} +  \frac{\sqrt{E_l - q_z}}{\sqrt{2 E_l}} \ |l(\breve q,2)\rangle_{\rm ch}\ .
\end{eqnarray}
(subindices $\mbox{ch}$ indicate that the chiral limit $m_l \rightarrow 0$ has been
taken). On the other hand, as mentioned above, for the LLL only one
polarization is available. Thus, the states associated with the spinors that
result from taking $m_l=0$ in Eqs.~(\ref{ULLL}) and~(\ref{VLLL}) are already helicity
eigenstates. We get
\begin{eqnarray}
\tilde \gamma_5 \ |l ((\breve q_{LLL})\rangle_{\rm ch} &=& s \ \mbox{sign}(q_z) \ |l (\breve q_{LLL})\rangle_{\rm ch} \ , \nonumber \\
\tilde \gamma_5 \ |\bar l ((\breve q_{LLL})\rangle_{\rm ch} &=& - s \ \mbox{sign}(q_z) \ |\bar l (\breve q_{LLL})\rangle_{\rm ch} \ .
\end{eqnarray}
This implies that for large enough magnetic fields ---such that only the LLL
is relevant and $m_l$ can be neglected--- a negatively charged lepton (like
the muon or the electron) is lefthanded if $B$ and $q_z$ are either both positive
or both negative, and it is righthanded otherwise.

We briefly consider now the pion eingenstates. Their angular momentum can be analyzed following similar steps
as before.
For the {\it canonical} orbital angular momentum we get
\begin{eqnarray}
\tilde l_z \ |\pi^\sigma(\breve p) \rangle \ = \  {l_z^{(\pi^\sigma)}} \ |\pi^\sigma(\breve p)\rangle \ ,
\end{eqnarray}
with
\begin{eqnarray}
{l_z^{(\pi^\sigma)}} \ = \ - s (\ell - \imath)\ ,
\label{jotapion}
\end{eqnarray}
where $s=\mbox{sign}(\sigma B)$. As in the case of charged leptons, the
{\it mechanical} angular momentum $L_z$ is {\it not} a conserved quantity,
thus, it is not diagonal in the basis of energy eigenstates.

\section{Calculation of the $\pi^-\to l\,\bar\nu_l$ decay width in the presence
of an external magnetic field using the symmetric gauge}

Let us analyze the decay width for the process $\pi^-\to l\,\bar\nu_l$, with
$l = \mu^-, e^-$, in the presence of a uniform magnetic field using the
symmetric gauge. For definiteness we will take $B>0\,$, i.e.\ $s=-1$.
Following the notation introduced in the previous section, the initial
charged pion state is determined by the quantum numbers $\breve p = (\ell,
\imath,p_z)$, the associated energy being $E_{\pi^-} =
\sqrt{m_{\pi^-}^2+(2\ell+1)B_e + p_z^2}$. The quantum numbers corresponding
to the outgoing lepton state are taken to be $\breve q = (n,\upsilon,q_z)$,
together with a polarization index $\tau$. In this case the energy is given
by $E_l = \sqrt{m_l^2+2nB_e + q_z^2}$. Finally, being electrically neutral,
the outgoing antineutrino is taken to be in the cylindrical basis discussed
in Appendix B. Thus, the associated quantum numbers are $\breve k =
(\jmath,k_\perp,k_z)$, where $\jmath$ is an integer while $k_\perp$ and
$k_z$ are real numbers, with $k_\perp >0$. The corresponding energy is
$E_{\bar\nu_l} = \sqrt{k_\perp^2+ k_z^2}$.

\subsection{The decay amplitude}

Using the notation introduced in the previous section, the transition
matrix element for the process we are interested in  is given by
\begin{eqnarray}
\langle\, l(\breve q , \tau)\, \bar \nu_l(\breve k, R ) | \mathcal{L}_W |
\pi^-(\breve p)\,\rangle & = &  - \frac{G_F}{\sqrt{2}}\, \cos \theta_c \times
\nonumber \\
& & \hspace{-1.7cm}  \int d^4x \, H^{-,\mu}_L(x,\breve p)
\ \bar  U^-_l(x,\breve {q}, \tau) \, \gamma_\mu \, (1-\gamma_5) \,
V_{\nu_l}(x,\breve k, R)\ ,
\end{eqnarray}
where $H^{-,\mu}_L(x,\breve p)$ stands for the matrix element of the hadronic
current,
\begin{equation}
H^{-,\mu}_{L}(x,\breve p) \ = \
H_V^{-,\mu} (x,\breve p) - H_A^{-,\mu} (x,\breve p) \ = \
\langle 0| \bar \psi_u(x)\, \gamma^\mu (1-\gamma_5)\, \psi_d(x)| \pi^-(\breve p) \rangle\ .
\label{hadronicme}
\end{equation}
Following the definitions and conventions of Ref.~\cite{Coppola:2018ygv}, the hadronic matrix elements can be
parametrized as
\begin{alignat}{2}
H^{-,\epsilon}_{\parallel,L} \ &= \ H^{-,0}_{L} +\epsilon H^{-,3}_{L} & \ &= \
\sqrt{2} \left(\epsilon f^{(V)}_{\pi^-} - f^{(A1)}_{\pi^-} \right)   \left( {\cal D}^0 +\epsilon {\cal D}^3 \right) W^-_{\bar p}(x)\ , \\
H^{-,\epsilon}_{\perp,L} \ &= \ H^{-,1}_{L} + i \epsilon H^{-,2}_{L} & \ &= \
- \sqrt{2} \left(f^{(A1)}_{\pi^-} + \epsilon f^{(A2)}_{\pi^-} - f^{(A3)}_{\pi^-}
\right) \left( {\cal D}^1 + i \epsilon {\cal D}^2 \right) W^-_{\bar p}(x)\ .
\end{alignat}
Rewriting $H^{-,\mu}_{L}$ in terms of the parallel and perpendicular components one gets
\begin{eqnarray}
\gamma_\mu \, (1-\gamma_5) \ H^{-,\mu}_L(x,\breve p) \ = \
\left(
  \begin{array}{cccc}
    0 & 0 & 0 & 0 \\
    0 & 0 & 0 & 0 \\
    H^{-,+}_{\parallel,L} &  H^{-,-}_{\perp,L} &\  0 \ & \ 0 \ \\
    H^{-,+}_{\perp,L} & H^{-,-}_{\parallel,L} &\ 0 \ & \  0 \ \\
  \end{array}
\right)\ ,
\end{eqnarray}
with
\begin{align}
H^{-,\epsilon}_{\parallel,L} \ &= \ - i \sqrt{2}\ \left(\epsilon f^{(V)}_{\pi^-} -   f^{(A1)}_{\pi^-} \right) \left( E_{\pi^-} +\epsilon p_z \right) \ W^-_{\bar p}(x)\ , \\
H^{-,\epsilon}_{\perp,L} \ &= \ - \epsilon \sqrt{2}  \left(f^{(A1)}_{\pi^-} +
\epsilon f^{(A2)}_{\pi^-} - f^{(A3)}_{\pi^-}  \right) \sqrt{( 2\ell + 1 +
\epsilon) B_e} \ W^-_{\bar p +\epsilon}(x) \ ,
\end{align}
where $\bar p +\epsilon = (E_{\pi^-},\ell+\epsilon,\imath,p_z)$.

Using these expressions together with the explicit form of the charged
lepton and antineutrino spinors (see Sec.~II.B and Appendix B) we get
\begin{eqnarray}
\langle\, l(\breve q , \tau)\, \bar \nu_l(\breve k, R ) | \mathcal{L}_W |
\pi^-(\breve p)\,\rangle & = & (2\pi)^3 \, \delta(E_{\pi^-} - E_l - E_{\bar\nu_l})\, \delta(p_z - q_z -
k_z)\times
\nonumber \\
& & \delta_{\ell-\imath, n-\upsilon + \jmath - 1}\, {\cal M}(\breve p, \breve q, \breve k, \tau)\ ,
\label{amp}
\end{eqnarray}
where
\begin{align}
{\cal M}(\breve p, \breve q, \breve k, 1) \, &= \, -\, \sqrt{2}\,G_F \cos \theta_c
\dfrac{(-i)^{\jmath+1}}{\sqrt{E_l+m_l}}  \Big[
(E_l+m_l - q_z)\,  A(\breve p, \breve q, \breve k)
+ \sqrt{2 n B_e}\, B(\breve p, \breve q, \breve k) \Big]\ , \nonumber \\
& & \nonumber\\
{\cal M}(\breve p, \breve q, \breve k, 2) \, &= \, \sqrt{2} \, G_F \cos \theta_c
\dfrac{(-i)^{\jmath} }{\sqrt{E_l+m_l}}\,
 \Big[ \sqrt{2 n B_e}\,  A(\breve p, \breve q, \breve k)
+ (E_l+m_l + q_z) \, B(\breve p, \breve q, \breve k) \Big]\ .
\label{emes}
\end{align}
Here we have used the definitions
\begin{align}
A(\breve p, \breve q, \breve k) \ &= \ a_{\pi^-} (E_{\pi^-} + p_z) \ \sqrt{E_{\bar\nu_l}-k_z} \ I_1
  - d_{\pi^-} \sqrt{2 \ell B_e} \ \sqrt{E_{\bar\nu_l}+k_z} \ I_2 \ , \nonumber \\
B(\breve p, \breve q, \breve k) \ &= \ b_{\pi^-}  (E_{\pi^-}-p_z) \sqrt{E_{\bar\nu_l}+k_z} \ I_4 -
c_{\pi^-} \sqrt{2 (\ell+1) B_e } \ \sqrt{E_{\bar\nu_l}-k_z}\; I_3\ ,
\label{ab}
\end{align}
and
\begin{align}
& a_{\pi^-} \ = \ f^{(A1)}_{\pi^-} - f^{(V)}_{\pi^-}\ ,
& b_{\pi^-} \ &= \ f^{(A1)}_{\pi^-} + f^{(V)}_{\pi^-}\ , \nonumber \\
& c_{\pi^-} \ = \ f^{(A1)}_{\pi^-} + f^{(A2)}_{\pi^-} - f^{(A3)}_{\pi^-}\ ,
& d_{\pi^-} \ &= \ f^{(A1)}_{\pi^-} - f^{(A2)}_{\pi^-} - f^{(A3)}_{\pi^-} \ ,
\label{defabc}
\end{align}
while $I_i$, $i=1,\dots 4$ are radial integrals given by
\begin{align}
& I_1 \ = \ {\cal I}(\ell,\imath,n-1,\upsilon)\ ,
& I_2 \ &= \ {\cal I}(\ell-1,\imath,n-1,\upsilon)\ , \nonumber \\
& I_3 \ = \ {\cal I}(\ell+1,\imath,n,\upsilon)\ ,
& I_4 \ &= \ {\cal I}(\ell,\imath,n,\upsilon)\ ,
\end{align}
where
\begin{equation}
{\cal I}(\ell,\imath,n,\upsilon) \ = \ 2\pi \int_0^\infty d\rho\ \rho \ R_{\ell, \imath}(\rho) \
R_{n,\upsilon}(\rho) \ J_{(\ell-\imath)-(n-\upsilon)}(k_\perp \rho)\ .
\label{integrals}
\end{equation}
The evaluation of these integrals for arbitrary particle states is given in Appendix C.

Note that, due to the $\delta$ functions appearing in Eq.~(\ref{amp}), not
all the variables in the expressions given in Eqs.~(\ref{emes}) are independent.
While the first two $\delta$ functions
correspond to the conservation of the energy and the $z$ component of the
linear momentum, the last one leads to the relation $\ell - \imath =
n-\upsilon-1/2 + \jmath - 1/2$. Taking into account that for
antineutrinos (in our basis) one has $j_z^{(\bar \nu_l)} = \jmath -1/2$ (see
Appendix B), using Eqs.~(\ref{jotalep}) and (\ref{jotapion}), and recalling
that we are considering $s=-1$, this relation can be written as
$j_z^{(\pi^-)} = j_z^{(l)} + j_z^{(\bar \nu_l)}$. Therefore, the last
$\delta$ function in Eq.~(\ref{amp}) implies that, as expected, the $z$
component of the {\it canonical} total angular momentum is conserved in the
decay process when the symmetric gauge is used.

Now let us concentrate on the situation in which the decaying pion is in the
lowest energy state (LES). This corresponds to $\ell =0$ and $p_z =0$, i.e.,
$\breve p_{LES} = (0,\imath,0)$. In this case the radial integrals $I_i$ get
simplified. Introducing $x= k_\perp^2/(2 B_e)$, we get
\begin{align}
I_1^{(\ell=0)} \ &= \ 2\pi\;  (-1)^{n-1} \; \frac{x^{(n-1)/2}}{\sqrt{(n-1)!}} \;
F_{\imath,\upsilon}\left(x\right)\ ,
\nonumber \\
I_2^{(\ell=0)} \ &= \ 0\ , \nonumber \\
I_3^{(\ell=0)} \ &= \  2\pi\; (-1)^n \; \frac{ x^{(n-1)/2}}{\sqrt{n!}} \; (x-n) \; F_{\imath,\upsilon}\left(x\right)\ ,
\nonumber \\
I_4^{(\ell=0)} \ &= \ 2\pi\; (-1)^n \; \frac{ x^{n/2}}{\sqrt{n!}} \;
F_{\imath,\upsilon}\left(x\right)\ ,
\end{align}
where
\begin{equation}
F_{\imath,\upsilon}(x) \ = \ \sqrt{\frac{\imath!}{\upsilon!}} \;
(-1)^{\imath+\upsilon} \; x^{(\upsilon-\imath)/2} \; e^{-x}
L_\imath^{\upsilon-\imath}(x)\ .
\label{fff}
\end{equation}
The decay amplitudes simplify to
\begin{multline}
{\cal M}(\breve p_{LES}, \breve q, \breve k, 1) = G_F \cos \theta_c \ (-1)^n \, (-i)^{\jmath+1} \, 2\pi \
\sqrt{\dfrac{2(E_{\bar\nu_l}-k_z)}{E_l+m_l}}\
\dfrac{x^{(n-1)/2}}{\sqrt{n-1!}} \ F_{\imath,\upsilon}(x) \,\times \\
\left[ a_{\pi^-}  E_{\pi^-}  (E_l+m_l - q_z)
- b_{\pi^-}  E_{\pi^-} (E_{\bar\nu_l} + k_z) + c_{\pi^-} 2 B_e (x-n) \right]\ ,
\label{ampLLL1}
\end{multline}
\begin{multline}
{\cal M}(\breve p_{LES}, \breve q, \breve k, 2) = G_F \cos \theta_c \ (-1)^{n+1} \, (-i)^\jmath \, 2\pi \
\sqrt{\dfrac{E_{\bar\nu_l}-k_z}{B_e(E_l+m_l)}}\
\dfrac{x^{(n-1)/2}}{\sqrt{n!}} \ F_{\imath,\upsilon}(x) \,\times \\
\left\{ a_{\pi^-} E_{\pi^-}\, 2n B_e
- (E_l+m_l + q_z) \left[ b_{\pi^-} E_{\pi^-} (E_{\bar\nu_l}+k_z) - c_{\pi^-} \, 2 B_e (x-n) \right] \right\}\ .
\label{ampLLL2}
\end{multline}

It is interesting at this point to consider the situation in which the
magnetic field is large enough so that the outgoing charged lepton can only
be in the lowest Landau level. As mentioned in the previous section, only
one polarization state is allowed in this case. Since we are considering
$s=-1$, this corresponds to $\tau = 2$. We get
\begin{eqnarray}
{\cal M}(\breve p_{LES}, \breve q_{LLL}, \breve k) &=& G_F \cos \theta_c \,
(-i)^{\jmath} \, 2\pi \,
\sqrt{\frac{2}{E_l+m_l}} \; F_{\imath,\upsilon}(x) \times \nonumber \\
& &    (E_l+m_l + q_z) \sqrt{E_{\nu_l}+k_z}\, \Big[ b_{\pi^-} \, E_{\pi^-} - c_{\pi^-} \, (E_{\nu_l}-k_z) \Big]\ ,
\label{ampLLL}
\end{eqnarray}
with $\breve q_{LLL} = (0,\upsilon,q_z)$. We remark that, due to the
$\delta$ functions in Eq.~(\ref{amp}), the relations $q_z = - k_z$, $\jmath
= \upsilon - \imath + 1$ and $k_z = \pm \sqrt{(E_{\pi^-}^2 + 2B_e x -
m_l^2)^2 - 8 B_e E_{\pi^-}^2 x}/(2 E_{\pi^-})$ are satisfied. Therefore,
recalling that $x= k_\perp^2/(2 B_e)$, we see that for fixed $E_{\pi^-}$ and
given definite values of $\imath$ and $\upsilon$ the amplitude is a function
of $k_\perp$. Contrary to the claim in Ref.~\cite{Bali:2018sey}, we conclude
that by no means angular momentum conservation implies that $k_\perp$ has to
be zero. In fact, one should integrate over the full range of values of
$k_\perp$ from zero to infinity to calculate the total width.

A final observation concerns the situation in which $B_e \gg m_l^2$. In this
case, we can neglect the charged lepton mass in Eq.~(\ref{ampLLL}), obtaining
\begin{eqnarray}
{\cal M}(\breve p_{LES}, \breve q_{LLL}, \breve k)_{\rm ch} &=& G_F \cos \theta_c\ (-i)^{\jmath} \;  2\pi \,
\sqrt{2E_l} \ F_{\imath,\upsilon}(x) \times \nonumber \\
& &   \left[1 - \mbox{sign}(k_z)\right] \sqrt{E_{\nu_l}+k_z}\;
\Big[ b_{\pi^-} \, E_{\pi^-}   - c_{\pi^-} \ (E_{\nu_l}-k_z)  \Big]\ ,
\end{eqnarray}
where we have used that  in the present case $E_l = |k_z| $. As seen, while
for $k_z > 0$ the amplitude vanishes, for $k_z < 0$ in general {\it it does
not}. This can be understood in terms of helicity conservation. As discussed
in Sec.~II.C, in the  limit $m_l\rightarrow 0$ for a charged lepton in the
LLL we have (recall once again that we are considering $s=-1$)
\begin{equation}
\gamma_5 \ |l (\breve q_{LLL})\rangle_{\rm ch} = - \ \mbox{sign}(q_z) \ |l (\breve q_{LLL})\rangle_{\rm
ch}\ .
\end{equation}
Noting that $q_z = - k_z$, we see that for $k_z > 0$ the outgoing charged
lepton would be righthanded, which is forbidden by helicity conservation
since antineutrinos are always righthanded. This is very different from what
happens in the absence of a magnetic field. For $B=0$, helicity conservation
implies that the total decay amplitude of a pion at rest must vanish as
$m_l\rightarrow 0$. At large magnetic field, however, it only implies that
the projection of the antineutrino momentum in the direction of $\vec B$
must be opposite to $\vec B$.

\subsection{Decay width}

On general grounds, the decay width for the process we are interested in
is given by
\begin{equation}
\Gamma_l^-(B)\ = \ \lim_{L,\,T\rightarrow\infty}  \sum_{\tau=1,2} \,\sum_{n,\upsilon,\jmath=0}^\infty
\int \!\frac{dq_z}{(2\pi)^3 2E_l} \dfrac{dk_z \ dk_\perp \ k_\perp}{(2\pi)^2 2E_{\bar\nu_l}}
\frac{|\langle\, l(\breve q ,\tau)\, \bar \nu_l( \breve k, R)
| \mathcal{L}_W |
\pi^-(\breve p)\,\rangle |^2}{2 (2\pi)^2 E_{\pi^-} L \, T}\ , \label{uno}
\end{equation}
where $T$ and $L$ are the time interval and length on the
$z$-axis in which the interaction is active. At the end of the
calculation, the limit $L,T\to \infty$ should be taken.
Using Eq.~(\ref{amp}) we get
\begin{eqnarray}
\Gamma_l^-(B)\, &=& \, \frac{1}{16\pi E_{\pi^-}}
\sum_{n,\upsilon,\jmath=0}^\infty \,\int \frac{dq_z \ dk_z \ dk_\perp \ k_\perp}{(2\pi)^2 E_l \,E_{\bar\nu_l}}
\times \nonumber \\
&& \, \delta(E_{\pi^-} - E_l - E_{\bar\nu_l}) \ \delta(p_z - q_z - k_z) \
\delta_{\ell-\imath, n-\upsilon-1 + \jmath} \
\overline{\big|{\cal M}_{\pi^-\to\, l\,\bar\nu_l}\big|^2}\ ,
\label{gamgen}
\end{eqnarray}
where
\begin{equation}
\overline{\big|{\cal M}_{\pi^-\to\, l\,\bar\nu_l}\big|^2} \ \equiv \
\sum_{\tau=1,2}\Big|{\cal M}(\breve p, \breve q, \breve k, \tau)\Big|^2
\label{ampa}
\end{equation}
and the amplitudes for $\tau =1,2$ are given in Eq.~\eqref{emes}.

Now, as it is usually done, we concentrate on the situation in which the
decaying pion is in the lowest energy state. This corresponds to $\ell =0$
and $p_z =0$. Moreover, as it will be shown below, the decay width will not
depend on the value of $\imath$. The expression in Eq.~(\ref{gamgen}) can be
worked out, leading to
\begin{equation}
\Gamma_{l}^{-}(B) \ = \ \dfrac{1}{16\pi E_{\pi^-}^2}
\sum_{n=0}^{n^{\rm max}}
\,\int_0^{k_\perp^{\rm max}}\dfrac{dk_{\perp} k_\perp}{2\pi}
\,\int\dfrac{dk_z}{2\pi \rule{0cm}{0.4cm}\bar{k_z}}\,
\big[\delta(k_z-\bar{k_z})+\delta(k_z+\bar{k_z})\big]\,
\sum_{\upsilon=0}^\infty \overline{\big|{\cal M}_{\pi^-\to\, l\,\bar\nu_l}\big|^2}\ ,
\label{gamgen-1}
\end{equation}
where we have used the definitions
\begin{align}
n^{\rm max} \ &= \ \frac{m_{\pi^-}^{2}-m_{l}^{2}+B_e}{2B_{e}}\ ,
\label{nmax} \\
k_\perp^{\rm max} \ &= \ E_{\pi^-}-\sqrt{m_l^2+2nB_{e}} \ , \\
\bar{k_z} \ &= \ \dfrac{1}{2E_{\pi^-}}
\sqrt{\left[E_{\pi^-}^2-2B_e(n-x)-m_l^2\right]^2-8B_e E_{\pi^-}^2 x}\ .
\label{k3}
\end{align}
In the amplitudes we must take $q_z=-k_z$ and $\jmath =
\upsilon+1-n-\imath$. Using Eqs.~(\ref{ampLLL1}) and (\ref{ampLLL2}) we get
\begin{equation}
\int\frac{dk_z}{2\pi}
\big[\delta(k_z-\bar{k_z})+\delta(k_z+\bar{k_z})\big]\,
\sum_{\upsilon=0}^\infty \overline{\big|{\cal M}_{\pi^-\to\, l\,\bar\nu_l}\big|^2} =
16 \pi \, G_F^2 \cos^2\theta_c\, \frac{x^{n-1}}{n!}\,
A_n(x) \sum_{\upsilon=0}^\infty F_{\imath,\upsilon}(x)^2\ ,
\label{wi1}
\end{equation}
where
\begin{align}
A^{(n)}_{\pi^-}(x) \ =\ &  \big[E_{\pi^-}^2 -2B_e(n-x)-m_l^2\big] \times \nonumber \\
& \bigg[\frac{m_l^2}{2}\,(n |a_{\pi^-}|^2 + x |b_{\pi^-}|^2) + B_e (n-x)
(n|a_{\pi^-} - c_{\pi^-} |^2+x|b_{\pi^-} - c_{\pi^-}|^2)\bigg]\, +
\nonumber \\
& 2B_e\,x\,
\bigg\{E_{\pi^-}^2\Big[n|a_{\pi^-} - b_{\pi^-} |^2-(n-x)|b_{\pi^-} - c_{\pi^-} |^2\Big]+(n-x)\,m_l^2|c_{\pi^-} |^2\,\bigg\}\, .
\end{align}
To proceed we have to evaluate the sum over the charged lepton quantum
number $v$ on the right-hand side of Eq.~(\ref{wi1}). As shown in Appendix D, one gets
\begin{equation}
\sum_{\upsilon=0}^\infty F_{\imath,\upsilon}(x)^2 \ = \ e^{-x}\ .
\label{iden}
\end{equation}
Using this result we arrive at the final expression for the $\pi^-\to l\,\bar\nu_l$
decay width, namely
\begin{equation}
\Gamma_l^-(B) \ = \
\frac{G_F^2\cos^2\theta_c}{2\pi\,E_{\pi^-}^2}\, B_e\,\sum_{n=0}^{n_{\rm
max}}\int_0^{x_{\rm max}}dx\  \frac{1}{\rule{0cm}{0.4cm}\bar{k_z}}\;\frac{x^{n-1}}{n!}\
e^{-x} \;A^{(n)}_{\pi^-}(x)\ .
\label{gamgenfinal}
\end{equation}
Our result agrees exactly with Eq.~(52) of Ref.~\cite{Coppola:2018ygv},
where the calculation was carried out using the Landau gauge. This provides
an additional and explicit confirmation of the gauge independence of our
expression for the decay width. It should be noted that, since the sum in
Eq.~(\ref{iden}) turns out to be independent of $\imath$, the width does not
depend on the charged pion {\it canonical} angular momentum $j_z^{(\pi^-)}$.
This is to be expected due to the fact that, as mentioned in Sec.~III,
$j_z^{(\pi^-)}$ is a gauge dependent quantity.

\section{Summary and conclusions}

In this work we study the decay width $\pi^- \to l \, \bar{\nu}_l$ in the
presence of an arbitrary large uniform magnetic field. We use here the
symmetric gauge, as an alternative to a previous analysis carried out in
Ref.~\cite{Coppola:2018ygv} where the Landau gauge was considered. The usage
of the symmetric gauge has the advantage of allowing for a better
understanding of the consequences of the axial symmetry of the problem, as
well as for a better treatment of angular momenta. In our analysis we
introduce charged pion and lepton wave functions and spinors in this gauge,
using cylindrical coordinates. To study the conservation of angular
momentum, we define the \textit{canonical} ($j_z$) and \textit{mechanical}
($J_z$) total angular momentum operators. We find that, as expected, even
though $j_z$ is not a gauge covariant quantity, it turns out to be conserved
in the symmetric gauge. On the other hand, $J_z$ is shown to be gauge
covariant but it is not conserved in \textit{any} gauge. As shown by
explicit calculation [see Eq.(41)], the relevant matrix element for the
process $\pi^- \to l \,\bar{\nu}_l$ turns out to be proportional to a
Kronecker delta which, when expressed in terms of the $j_z$ values of the
different fields, clearly reflects the conservation of the
\textit{canonical} total angular momentum of the system. Using the symmetric
gauge we also obtain an explicit expression for the decay width $\pi^- \to l\,
\bar{\nu}_l$ for the case in which the decaying pion lies in its state of
minimum energy (i.e., in the lowest Landau level, with zero linear momentum
along the direction of the magnetic field). We show that the total width
does not depend on the charged pion {\it canonical} angular momentum
$j_z^{(\pi^-)}$, a fact that is to be expected since $j_z^{(\pi^-)}$ is a
gauge dependent quantity. Moreover, it is seen that the total width is
obtained after integrating over the perpendicular momenta of the outgoing
antineutrinos, $k_\perp$. This confirms that angular momentum conservation
does not imply, as claimed in Ref.~\cite{Bali:2018sey}, that the momentum of
the antineutrino has to be necessarily parallel to the magnetic field. In
fact, it turns out that ---as showed in Ref.~\cite{Coppola:2019idh} for
large magnetic fields--- the ratio $k_\perp/\sqrt{k_\perp^2+k_z^2}$ for
outgoing antineutrinos tends to be relatively large. As expected, the
derived expression for the integrated $\pi^- \to l \,\bar{\nu}_l$ width
coincides exactly with the one found using the Landau gauge in
Ref.~\cite{Coppola:2018ygv}. This provides an explicit test of the gauge
independence of this result.

It is worth noticing that Eq.~(\ref{gamgenfinal}) implies that for
finite magnetic fields the decay width does not vanish in the limit $m_l \to
0$, i.e., it does not exhibit the helicity suppression found in the case of
no external field. As shown in the present work, for a sufficiently large
magnetic field (so that the outgoing charged lepton has to be in the LLL,
and its mass can be neglected), helicity conservation only implies that the
projection of the antineutrino momentum in the direction of the magnetic
field should be antiparallel to the magnetic field. As a consequence,
for large values of $B$ the ratio $\Gamma_e/\Gamma_µ$ might change
dramatically with respect to the $B = 0$ value~\cite{Coppola:2019idh}. This
could be interesting e.g.~regarding the expected flavor composition of
neutrino fluxes coming from the cores of magnetars and other stellar
objects. In addition, for large $B$ the angular distribution of outgoing
antineutrinos is expected to be highly anisotropic, showing a significant
suppression in the direction of the external field.

\begin{acknowledgments}

This work has been supported in part by Consejo Nacional de Investigaciones
Cient\'ificas y T\'ecnicas and Agencia Nacional de Promoci\'on Cient\'ifica
y Tecnol\'ogica (Argentina), under Grants No.~PIP17-700 and No.
PICT17-03-0571, respectively; by the National University of La Plata
(Argentina), Project No.~X824; by the Ministerio de Econom\'ia y
Competitividad (Spain), under Contract No.~FPA2016-77177-C2-1-P; and by the
Centro de Excelencia Severo Ochoa Programme, Grant No.~SEV-2014-0398.

\end{acknowledgments}

\section*{Appendix A: Alternative representation of magnetized fermion spinors
in the symmetric gauge}

\addtocounter{section}{1}
\setcounter{equation}{0}
\renewcommand{\theequation}{A\arabic{equation}}

For completeness, in this appendix we present an alternative way to express
the charged fermion spinors. This form follows quite closely the Ritus
representation notation often used in the Landau gauge (see e.g.\ Appendix~A.3 of
Ref.~\cite{Coppola:2018ygv}). In fact, the spinors in
Eq.~(\ref{fermionfieldBpart}) can also be written as
\begin{align}
U_l^s\left(x,\breve{q},\tau\right)  \ &= \ \mathbb{N}^{s}_{\bar q}(x) \ u_l^s\left(\breve{q},\tau\right)\ ,
\nonumber\\
V_l^{-s}\left(x,\breve{q},\tau\right) \ &= \ \mathbb{\tilde{N}}^{-s}_{\bar q}(x) \
v_l^{-s}\left(\breve{q},\tau\right)\ ,
\label{UVlept}
\end{align}
where $\bar q = (q^0, \breve{q})$, with $q^0 = E_l$. The spinors $u_l^s$ and
$v_l^{-s}$ are given in the Weyl basis by
\begin{align}
u_l^s\left(\breve {q},\tau\right)\ &=\
\frac{1}{\sqrt{2(E_l+m_l)}}
\left( \begin{array}{c}
  ( E_l + m_l + s \sqrt{2 n B_e}\ \tau_2 - q_z \tau_3) \phi^{(\tau)} \\
  ( E_l + m_l - s \sqrt{2 n B_e}\ \tau_2 + q_z \tau_3) \phi^{(\tau)}\\
\end{array}
\right)\ ,
\label{uspindos}
\\
v_l^{-s}\left(\breve {q},\tau\right)\ &=\
\frac{1}{\sqrt{2(E_l+m_l)}}
\left( \begin{array}{c}
  ( E_l + m_l - s \sqrt{2 n B_e} \ \tau_2 - q_z \tau_3) \tilde \phi^{(\tau)} \\
  -( E_l + m_l + s \sqrt{2 n B_e} \ \tau_2 + q_z \tau_3) \tilde \phi^{(\tau)}\\
\end{array}
\right)\ ,
\label{vspindos}
\end{align}
where $\tau_i$ are the Pauli matrices while $\phi^{(1)}{}^\dagger = -\tilde \phi^{(2)}{}^\dagger = (1,0)$ and
$\phi^{(2)}{}^\dagger = \tilde \phi^{(1)}{}^\dagger = (0,1)$. They satisfy the
relations
\begin{align}
\sum\limits_{\tau=1,2} u_l^s(\breve{q}, \tau) \, \bar u_l^s(\breve{q}, \tau)\ &=\  \rlap/\hat q_s + m_l\ ,
\nonumber\\
\sum\limits_{\tau=1,2} v_l^{-s}(\breve{q}, \tau) \, \bar v_l^{-s}(\breve{q}, \tau)\ &=\ \rlap/\hat q_{-s} -
m_l\ ,
\label{sumspinl}
\end{align}
where $\hat q_s^{\,\mu} = (E_l, 0, - s \sqrt{ 2 n B_e}, q_z)$. In
Eq.~(\ref{UVlept}), $\mathbb{N}^{s}_{\bar q}(x)$ and
$\mathbb{\tilde{N}}^{-s}_{\bar q}(x)$ are the symmetric gauge equivalents of
the Landau gauge Ritus functions. They are solutions of the eigenvalue
equation
\begin{equation}
\rlap/\!{\cal D}^2 \ \mathbb{N}^{s}_{\bar q} (x) \ = \
- \left[ E_l^2 - 2 n B_e - q_z^2 \right] \mathbb{N}^{s}_{\bar q}(x)\ ,
\label{ecautov}
\end{equation}
where $\rlap/\!{\cal D} =  \rlap/\partial - i s B_e \, (r_x \gamma^2 - r_y \gamma^1)$. The
explicit form of these functions is
\begin{equation}
\mathbb{N}^{s}_{\bar q} (x)\ = \ \sum_{\lambda=\pm} N^{s}_{\bar q,\lambda}(x)\,\Delta^\lambda\ ,
\qquad\qquad
\mathbb{\tilde{N}}^{-s}_{\bar q}(x) \ = \ \sum_{\lambda=\pm} N^{-s}_{\bar q,-\lambda}(x)^\ast
\Delta^{\lambda}\ ,
\label{ep}
\end{equation}
where $\Delta^{\pm} = (1\pm i \gamma^1 \gamma^2)/2$ and
\begin{equation}
N^{s}_{\bar q,\lambda}(x)  \ = \
W_{(E_l,n_{s\lambda},\upsilon,q_z)}^s(x)\ ,
\label{autofuncion}
\end{equation}
$W_{\bar q}^s(x)$ being given by Eq.~(\ref{efes}). Here the non-negative integer index
$n_{s\lambda}$ is related to the quantum number $n$ by $n_{s\pm}  \ = \ n - (1\mp s)/2$.

\section*{Appendix B: Neutrino field in a cylindrical basis}

\addtocounter{section}{1}
\setcounter{equation}{0}
\renewcommand{\theequation}{B\arabic{equation}}

It is usual to write the neutrino field as a linear combination of operators
of well defined linear momentum $\vec k$. However, this is not very
convenient for the purpose of the present work. As mentioned in the main
text, we are interested in dealing with the decay of charged pions in the
presence of an external field using the symmetric gauge. Thus, as in the
case of charged pions and leptons, it is more convenient to introduce an
expansion of the neutrino field using cylindrical coordinates $\rho$, $\phi$
and $z$, where $z$ is a spacial axis
parallel to the magnetic field. We can then expand the usual plane wave
functions in terms of eigenfunctions of the $z$ component of the orbital
momentum, $l_z$. Next, we can couple these wave functions to the eigenstates
of $S_z$ ($z$ component of the spin operator), and write the neutrino
field in terms of operators with ``good'' total angular momentum $j_z = l_z
+ S_z$. The resulting expansion reads
\begin{equation}
\psi_{\nu_l}(x) \ = \
\sum_{\jmath=-\infty}^\infty \int \! \frac{dk_z \, dk_\perp \, k_\perp}{(2\pi)^2 \, 2 E_{\nu_l}}
\left[ b(\breve k, L) \ U_{\nu_l}(x,\breve k,L) + d(\breve k, R)^\dagger  \ V_{\nu_l}(x,\breve k,R)
\right]\ ,
\label{neutrinofield}
\end{equation}
where  $\breve k = (\jmath, k_\perp, k_z)$ and $E_{\nu_l} = E_{\bar\nu_l} =
\sqrt{k_\perp^2 + k_z^2}\,$. Here, $\jmath$ is an integer number, while
$k_\perp$ and $k_z$ are real numbers, with $k_\perp > 0$. In the Weyl basis,
the spinors $U_{\nu_l}$ and $ V_{\nu_l}$ are given by
\begin{align}
U_{\nu_l}\left(x,\breve k,L\right)\ &=\
-\, i^{\jmath} \ e^{-i(E_{\bar\nu_l} t - k_z r_z)} \ e^{-i \jmath \; \phi}
\left( \!\!\! \begin{array}{c}
      \sqrt{ E_{\bar\nu_l} - k_z} \ J_{\jmath}(k_\perp \rho) \\
      i \sqrt{ E_{\bar\nu_l} + k_z}  \ e^{i \phi} \ J_{\jmath-1}(k_\perp \rho) \\ 0\\ 0
\end{array}
\!\!\! \right)\ ,
\label{Uspindos}
\\ && \nonumber
\\
V_{\nu_l}\left(x,\breve k,R\right)\ &=\
- (-i)^{\jmath} \ e^{i(E_{\bar\nu_l} t - k_z r_z)} \ e^{-i \jmath \; \phi}
\left( \!\!\! \begin{array}{c}
      \sqrt{ E_{\bar\nu_l} - k_z} \ J_{\jmath}(k_\perp \rho) \\
   -i \sqrt{ E_{\bar\nu_l} + k_z} \ e^{i\phi} \ J_{\jmath-1}(k_\perp \rho)\\ 0\\ 0
\end{array}
\!\!\! \right)\ .
\label{Vspindos}
\end{align}
Note that, as it is clear from the explicit form of the spinors, in the expansion
we have already taken into account that neutrinos (antineutrinos) are lefthanded (righthanded).
The creation and annihilation operators satisfy
\begin{eqnarray}
\left\{b(\breve{k},L), b(\breve{k}\,',L)^\dagger\right\} &=&
\left\{d(\breve{k},R), d(\breve{k}\,',R)^\dagger\right\} =  \,
2 E_{\bar\nu_l}\, (2\pi)^2\,  \delta_{\breve{k},\breve{k}^{\prime}}\ ,
\nonumber \\
\left\{b(\breve{k},L), d(\breve{k}\,',R)^\dagger\right\} &=&
\left\{d(\breve{k},L)^\dagger, b(\breve{k}\,',R)^\dagger\right\}\ = \, 0\ ,
\label{conferB}
\end{eqnarray}
where
\begin{eqnarray}
\delta_{\breve{k},\breve{k}^{\prime}} = \delta_{\jmath \jmath'}\, \frac{\delta(k_\perp - k_\perp')}{k_\perp} \, \delta(k_z - k'_z)\ .
\end{eqnarray}
It can be seen that the spinors in Eqs.~(\ref{Uspindos}) and
(\ref{Vspindos}) satisfy the orthogonality relations
\begin{alignat}{2}
\int\! d^3 r  \, U(x,\breve{k}, L)^\dagger\ U(x,\breve{k}', L) &=
\int\! d^3 r \, V(x,\breve{k}, R)^\dagger \ {V}(x,\breve{k}', R) & &=
2 E_{\bar\nu_l} \, (2\pi)^2 \delta_{\breve{k},\breve{k}^{\prime}}\ ,
\\
\int\! d^3 r  \, U(x,\breve{k}, L)^\dagger\ V(x,\breve{k}', R) &=
\int\! d^3 r \, V(x,\breve{k}, R)^\dagger \ {U}(x,\breve{k}', L) & &= 0 \ .
\end{alignat}
Using methods similar to those mentioned in Sec.~II.C it can be shown that,
given a set of quantum numbers $(\jmath, k_\perp, k_z)$, the eigenvalue of
the total angular angular momentum operator $\tilde j_z$ acting on a neutrino
state is $j_z^{(\nu_l)} = - (\jmath-1/2)$, while for an antineutrino state
one has $j_z^{(\bar \nu_l)} = \jmath-1/2$.

\section*{Appendix C: The radial integral}

\addtocounter{section}{1}
\setcounter{equation}{0}
\renewcommand{\theequation}{C\arabic{equation}}

In this appendix we quote the result for the radial integral appearing in Eq.~(\ref{integrals}).
It can be calculated using the relation (see Eq.~(5) of Ref.~\cite{Kolbig:1995kw})
\begin{multline}
\int dx \ x^{\nu+1} \ e^{-\alpha x^2} L_m^{\nu-\sigma}(\alpha x^2)\ L_n^{\sigma}(\alpha x^2)
 \ J_\nu(x y) \ = \ \dfrac{(-1)^{m+n}}{2 \alpha} \;\left(\dfrac{y}{2 \alpha}\right)^\nu
 \exp\Big(-\dfrac{y^2}{4\alpha}\Big)\,\times \\
L_m^{\sigma-m+n}\left(\frac{y^2}{4\alpha}\right)\ L_n^{\nu-\sigma+m-n}\left(\frac{y^2}{4\alpha}\right)
\ \ [y > 0, \mbox{Re}\; \alpha >0 , \mbox{Re}\; \nu > -1] \ ,
\end{multline}
together with (see Eq.~(3.6.2) of Ref.~\cite{doman})
\begin{equation}
L_\upsilon^{n-\upsilon}(x) \ = \ \dfrac{n!}{\upsilon!}(-x)^{\upsilon-n} L_n^{\upsilon-n}(x)\ ,
\label{lnegative}
\end{equation}
which is valid for all the values of $n$ and $\upsilon$ we are interested in. We get
\begin{equation}
{\cal I}(\ell,\imath,n,\upsilon) \ = \ 2\pi \; (-1)^{\imath+\upsilon} \sqrt{\frac{\imath!\, n!}{\ell!\,
\upsilon !}}  \left(\frac{k_\perp^2}{2
B_e}\right)^{\frac{(\ell-\imath)-(n-\upsilon)}{2}}
e^{-\frac{k_\perp^2}{2 B_e}}\ L_{\imath}^{\upsilon-\imath} \left(
\frac{k_\perp^2}{2 B_e} \right) \  L_n^{\ell-n} \left(
\frac{k_\perp^2}{2 B_e} \right)\ .
\end{equation}

\section*{Appendix D: Sum over the quantum number $\upsilon$ of the outgoing charged leptons}

\addtocounter{section}{1}
\setcounter{equation}{0}
\renewcommand{\theequation}{D\arabic{equation}}

In this appendix we will prove the validity of Eq.~(\ref{iden}). From the
definition of $F_{\imath,\upsilon}(x)$ in Eq.~(\ref{fff}) we have
\begin{equation}
\sum_{\upsilon=0}^\infty F_{\imath,\upsilon}(x)^2 \ = \ e^{-2x} S_\imath(x)\ ,
\label{uno}
\end{equation}
where
\begin{equation}
S_\imath(x) \ = \ \sum_{\upsilon=0}^\infty \, \frac{\imath !}{\upsilon!} \ x^{\upsilon-\imath} \,
L_\imath^{\upsilon - \imath}(x)^2\ .
\label{sx}
\end{equation}
We want to show that $S_\imath(x) = e^x$, for all $\imath\, \epsilon\, {\bf
N}$. For $\imath = 0$ one has $L_0^\upsilon(x) = 1$, therefore the relation
is clearly satisfied. For $\imath = 1$, using $L_1^{\upsilon-1}(x) =
\upsilon - x$, one has
\begin{eqnarray}
S_1(x) & = &
x \; + \; \sum_{\upsilon=1}^\infty \, \frac{x^{\upsilon-1}}{\upsilon!} \,
(\upsilon - x)^2 \nonumber \\
& = &
x \; + \; \sum_{\upsilon=1}^\infty \, \Big[
\frac{x^{\upsilon+1}}{\upsilon!} \, + \,
\frac{\upsilon\,x^{\upsilon-1}}{(\upsilon -1)!}\;-\,
\frac{2\,x^{\upsilon}}{(\upsilon-1)!}\Big] \ ,
\end{eqnarray}
from which $S_1(x) = e^x$ easily follows. For $\imath \geq 2$, let us
calculate the derivative of $S_\imath(x)$ with respect to $x$.
Using Eq.~(\ref{lnegative}) one can write
\begin{equation}
S_\imath(x) \ = \
\sum_{\upsilon=0}^{\imath-1} \, \frac{\upsilon !}{\imath!} \ x^{\imath-\upsilon} \,
L_\upsilon^{\imath - \upsilon}(x)^2 \ + \
\sum_{\upsilon=\imath}^\infty \, \frac{\imath !}{\upsilon!} \ x^{\upsilon-\imath} \,
L_\imath^{\upsilon - \imath}(x)^2 \ .
\label{sx2}
\end{equation}
In this way, using the general relation
\begin{equation}
\frac{dL_n^\alpha(x)}{dx} \ = \ -\,L_{n-1}^{\alpha+1}(x)\ ,
\end{equation}
one has
\begin{equation}
\frac{dS_\imath(x)}{dx} \ = \ \frac{x^{\imath-1}}{(\imath-1)!}
+ S^{(1)}_\imath(x) - 2\,L_\imath(x)\,L^1_{\imath -1}(x) +
S^{(2)}_\imath(x) \ ,
\label{ds}
\end{equation}
where
\begin{eqnarray}
S^{(1)}_\imath(x) & = & \sum_{\upsilon=1}^{\imath-1} \, \frac{\upsilon!}{\imath!} \Big[
(\imath - \upsilon) \ x^{\imath-\upsilon-1} \, L_\upsilon^{\imath - \upsilon}(x)^2 \ - \
2 \, x^{\imath-\upsilon} \, L_\upsilon^{\imath - \upsilon}(x)\,
L_{\upsilon-1}^{\imath - \upsilon+1}(x) \Big]\ , \nonumber \\
S^{(2)}_\imath(x) & = & \sum_{\upsilon=\imath+1}^\infty \, \frac{\imath!}{\upsilon!} \Big[
(\upsilon - \imath) \ x^{\upsilon-\imath-1} \, L_\imath^{\upsilon-\imath}(x)^2 \ - \
2\, x^{\upsilon-\imath} \, L_\imath^{\upsilon-\imath}(x)\, L_{\imath-1}^{\upsilon-\imath+1}(x) \Big] \ .
\label{sumas}
\end{eqnarray}

The sums in Eqs.~(\ref{sumas}) can be worked out using the relations
\begin{equation}
\alpha\,L_n^\alpha(x)\ = \ x\,L_n^{\alpha+1}(x)\,+\,(n+1)\,L_{n+1}^{\alpha-1}(x)
\end{equation}
and
\begin{equation}
L_{n+1}^\alpha(x) \ = \ L_{n+1}^{\alpha+1}(x)-L_n^{\alpha+1}(x)\ .
\label{rel3}
\end{equation}
For the first sum one has
\begin{eqnarray}
S^{(1)}_\imath(x) & = & \sum_{\upsilon=1}^{\imath-1} \, \Big[
(\imath - \upsilon)\,L_\upsilon^{\imath
- \upsilon}(x) \, - \, 2x\, L_{\upsilon-1}^{\imath -
\upsilon+1}(x)\Big] \, \frac{\upsilon !}{\imath!}\,
x^{\imath-\upsilon-1} \, L_\upsilon^{\imath
- \upsilon}(x) \\
& = & \sum_{\upsilon=1}^{\imath-1} \, \Big[\frac{\upsilon !}{\imath!}\,
x^{\imath-\upsilon} \, L_\upsilon^{\imath
- \upsilon}(x)^2\, +\, \frac{(\upsilon+1)!}{\imath!}\,
x^{\imath-\upsilon-1} \, L_{\upsilon+1}^{\imath
- \upsilon-1}(x)\,L_\upsilon^{\imath
- \upsilon}(x) \nonumber \\
& & -\, \frac{\upsilon!}{\imath!}\,
x^{\imath-\upsilon} \, L_{\upsilon}^{\imath
- \upsilon}(x)\,L_{\upsilon-1}^{\imath
- \upsilon+1}(x)\Big] \\
& = & L_\imath(x)\,L_{\imath-1}^1(x)\,-\,(\imath-x)\,\frac{x^{\imath-1}}{\imath!}
\, + \,
\sum_{\upsilon=1}^{\imath-1} \, \frac{\upsilon !}{\imath!}\,
x^{\imath-\upsilon} \, L_\upsilon^{\imath
- \upsilon}(x)^2\ ,
\label{s1}
\end{eqnarray}
where we have used $L_0^\imath(x) = 1$, $L_1^{\imath-1}(x) =
\imath - x$. The second sum is given by
\begin{eqnarray}
S^{(2)}_\imath(x) & = & \sum_{\upsilon = \imath+1}^\infty \, \Big[\frac{\imath
!}{\upsilon!}\, x^{\upsilon-\imath} \, L_\imath^{\upsilon - \imath}(x)^2\,
+\, \frac{(\imath+1)!}{\upsilon!}\, x^{\upsilon-\imath-1} \,
L_{\imath+1}^{\upsilon-\imath-1}(x)\,L_\imath^{\upsilon-\imath}(x)\nonumber \\
& & -\, \frac{\imath!}{\upsilon!}\,
x^{\upsilon-\imath} \, L_{\imath}^{\upsilon-\imath}(x)\,
L_{\imath-1}^{\upsilon-\imath+1}(x)\Big]\nonumber \\
& = & \sum_{\upsilon = \imath+1}^\infty \, \frac{\imath
!}{\upsilon!}\, x^{\upsilon-\imath} \, L_\imath^{\upsilon - \imath}(x)^2\,
+\, L_{\imath+1}(x)\,L_\imath^1(x) \nonumber \\
& & +\, \sum_{\upsilon = \imath+2}^\infty \, \frac{\imath
!}{\upsilon!}\, x^{\upsilon-\imath-1} \Big[
(\imath+1)\,L_{\imath+1}^{\upsilon-\imath-1}(x)\,L_\imath^{\upsilon-\imath}(x)
- \upsilon\, L_{\imath-1}^{\upsilon-\imath}(x)\,
L_\imath^{\upsilon-\imath-1}(x)\Big]\ .
\label{s2}
\end{eqnarray}
Here one can use the relation
\begin{equation}
(n+1)\,L_{n+1}^\alpha(x)\ = \ (n+\alpha+1)\,L_n^\alpha(x)\,-\,x\,L_n^{\alpha+1}(x)
\label{rel2}
\end{equation}
together with Eq.~(\ref{rel3}) to obtain
\begin{equation}
(\imath+1)\,L_{\imath+1}^{\upsilon-\imath-1}(x)\,L_\imath^{\upsilon-\imath}(x)
- \upsilon\, L_{\imath-1}^{\upsilon-\imath}(x)\,
L_\imath^{\upsilon-\imath-1}(x) \ = \
\upsilon \,L_{\imath}^{\upsilon-\imath-1}(x)^2\,-\,x\,
L_\imath^{\upsilon-\imath}(x)^2 \ .
\end{equation}
Replacing into Eq.~(\ref{s2}) and performing an adequate change in the index
of the sum it is easy to arrive at
\begin{equation}
S^{(2)}_\imath(x) \ = \ \sum_{\upsilon = \imath+1}^\infty \, \frac{\imath
!}{\upsilon!}\, x^{\upsilon-\imath} \, L_\imath^{\upsilon - \imath}(x)^2\,
+\, L_{\imath+1}(x)\,L_\imath^1(x) \,+\, \frac{x}{\imath+1}\,
L_\imath^1(x)^2\ .
\end{equation}
Now replacing the expressions for $S^{(1)}_\imath(x)$ and
$S^{(2)}_\imath(x)$ in Eqs.~(\ref{s1}) and (\ref{s2}) into Eq.~(\ref{ds}),
and using Eq.~(\ref{rel3}), one obtains
\begin{eqnarray}
\frac{dS_\imath(x)}{dx} & = &
\sum_{\upsilon=0}^{\imath-1} \, \frac{\upsilon !}{\imath!} \ x^{\imath-\upsilon} \,
L_\upsilon^{\imath - \upsilon}(x)^2 \ + \
\sum_{\upsilon=\imath}^\infty \, \frac{\imath !}{\upsilon!} \ x^{\upsilon-\imath} \,
L_\imath^{\upsilon - \imath}(x)^2 \nonumber \\
& & +\, L_\imath^1(x)\,\Big[
L_{\imath+1}(x)-L_\imath(x)+\frac{x}{\imath+1}\,
L_\imath^1(x)\Big] \ ,
\end{eqnarray}
where the last term on the right-hand side vanishes, according to Eq.~(\ref{rel2}).
Thus, taking into account the definition of $S_\imath(x)$ in Eq.~(\ref{sx2}),
it is seen that $dS_\imath(x)/dx = S_\imath(x)$. Since, in addition,
for all $\imath$ one has $S_\imath(0)=L_\imath(0)^2=1$, one obtains
$S_\imath(x)=e^x$. From Eq.~(\ref{uno}) one has finally
\begin{equation}
\sum_{\upsilon=0}^\infty F_{\imath,\upsilon}(x)^2 \ = \ e^{-x}\ .
\nonumber
\end{equation}

\end{document}